# Spatial structure of disordered media: Unravelling the mechanical significance of disorder in granular materials

K.P.Krishnaraj

*Department of Chemical Engineering, Indian Institute of Science, Bangalore, India*

e-mail: krishnarajkp@gmail.com

**Abstract**

For two decades now, the importance of microstructure in the mechanical behaviour of a large collection of grains has been a topic of intense research[1-16]. Many approaches have been developed to study the microstructure in granular materials, based on local or mesoscopic measures[3,6,8-11] or using ideas from diverse fields like percolation theory[7,12], complex networks[7,13,14] and persistent homology[15,16]. However, we do not fully understand the role of microstructure in the mechanical behaviour of simple and commonplace packings like a pile of sand[2,17-26] or box of grains[1,2,7,17,27,28]. Here, by characterizing the spatial variation of local microstructure, we uncover intriguing large-scale spatial patterns in particle packings deposited under the influence of gravity. We detail the emergence of these patterns, and provide a unified fundamental explanation of classic and puzzling mechanical behaviours of granular materials like central stress minimum[2,17-26] and Janssen effect[7,27,28]. Further, we show the striking dependence of spatial structure on the history of preparation, size distribution of particles and boundary conditions. Our study reveals the existence of global spatial structure in a locally disordered media, and elucidates its significance in the mechanical behaviour of granular materials.

**Introduction**

Particulate media are a common form of matter that we encounter in our daily lives. Granular materials are important and more familiar representative of particulate media, considering their prevalence in nature and relevance in processing industries[1,2,17]. Also, force transmission in granular media is strikingly similar to other particulate media like foams[29], emulsions[30,31], gels[32,33], and even bacterial suspensions[34]. However, the mechanical behaviour of even the simplest of geometries like a pile[2,17-26] or a vertical column of grains[1,2,7,17,27,28] has proved to be challenging to model and, no general explanation is available[1,2,17]. In granular piles created by pouring grains from a narrow source like a funnel, the normal stress at the base exhibits a local minimum beneath the apex[2,17-26], in contrast to intuitive expectation. And, in a vertical column of grains filled by raining, the components of wall stress asymptotically saturate with depth from the free surface, commonly referred to as the Janssen effect[7,27,28]. Some studies suggest that the microstructure might be the cause of central stress minimum in granular piles[21-25,35-38], but the evidences are not conclusive. In a granular column though, the role of microstructure in Janssen effect is unclear and has not been studied in detail before[7,28]. Many explanations of these mechanical anomalies have been proposed[7,27,35-43], but a generally accepted consensus is yet to be reached[1,2,17,43]. Further, these intriguing mechanical behaviours have inspired numerous studies on the microstructure of granular materials[3-16].

Our current understanding of the microstructure of granular materials is based on local particle scale measures like the contact vector orientations[3,6] or mesoscopic structures made of particles[8-10] or force chain orientations[11,23,24] (also mesoscopic) or percolation analysis of the contact force network[7,12] or complex network measures[7,13,14] like communities[14] or topological invariants like Betti numbers[15,16]. In a recent study, the need and advantages of a long-ranged measure of microstructure over the existing local particle scale or mesoscopic approaches was shown[7]. A comprehensive and generally accepted description of the microstructure in granular materials is yet to be found. Hence, in modelling force transmission through granular materials, the details of the microstructure are not adequately represented. For example, in the widely studied $q$ model of force transmission[44], the effects of contact angles of particles are chosen from a common distribution without regard to the spatial position, and the model does not explain the central stress minimum in piles or Janssen effect in silos. However, anisotropic extensions of the $q$ model can predict the central stress minimum in granular piles[41]. In addition, some studies on the microstructure of granular piles show that, contact vector distributions considerably differ depending on the geometric region of the pile studied[25]. In point force experiments too, the spatial ordering of the packing was shown to influence force transmission[4,5]. Further, some continuum models that account for structural anisotropy based on intuitive ideas like arching do predict the central stress minimum[35-38]. Hence, the spatial dependence of microstructure in granular materials is important and is not well understood.

In this work, using a simple yet novel method to characterise the spatial variation of local microstructure, we reveal striking spatial patterns in granular packings under gravity, and provide a unified explanation to central stress minimum in piles and Janssen effect in silos. We show that, the spatial structure in granular piles surprisingly emerges from a nearly horizontal flow in the core region, in complete contrast to surface avalanches hypothesis of many previous studies[21-25,35-38,41,42]. Further, we show that, the spatial structure can vary dramatically depending on the method of deposition, boundary conditions and size distribution of the particles. The details of the simulation method, various systems studied and deposition procedure used for creating pile and silo packings are provided in Supplementary Note 1 & 2.

**Results and discussion**

First, we describe a method to qualitatively present the spatial pattern of the contact network. In packing's deposited only under the influence gravity, the weight is transmitted in a preferred direction i.e. the direction of gravity. Hence, we find the average orientation of contacts in the direction of gravity and their spatial distribution. From every particle centre, we define a vector, $\vec{R}$ based on the resultant direction of its contacts, here we only consider contact directions with a positive vertical component in the direction of gravity (see Fig. 1). For example, in Fig. 1, from particle A, a contact A-B is considered only if $\vec{n}_{ab} \cdot \vec{g} > 0$, where $\vec{n}_{ab}$ is the normalized contact vector and $\vec{g}$ is a unit vector in the direction of gravity. Thus, for every particle, a vector is assigned, and a vector field is obtained for the complete packing. The resulting vector field is also averaged over many independent realizations. Then, like the streamlines of a continuum velocity field, we draw lines that represent the local direction of the contact vector field at every spatial point. We refer to this representation of the contact network as contact lines of the packing.

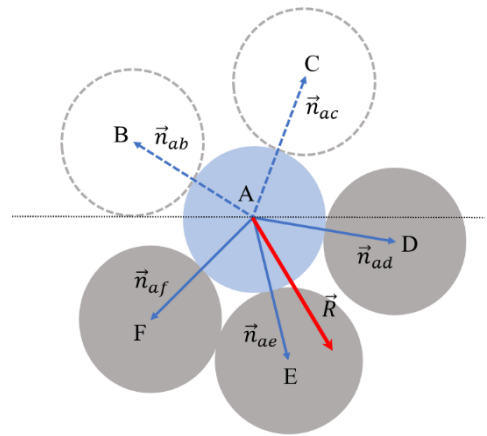

**Figure 1: Definition of $\vec{R}$.** For particle A (coloured in blue), only contacts vectors $\vec{n}_{ai}$, with $\vec{n}_{ai} \cdot \vec{g} > 0$ (shown in grey) are considered for finding $\vec{R}$ (shown in red), where $\vec{g}$ is the unit vector in the direction of gravity, and $i$ is the index of all particles in contact with A.

Now, we describe the spatial structure of pile and silo packings using contact lines. As shown in Fig. 2,3, emergence of large-scale spatial pattern in the contact network is clear in the contact line representation. In piles created by narrow source deposition method (Fig. 2a, b), along the vertical height of the pile, the contact lines are directed away from the centre. In contrast, the contact lines in piles created by raining particles are aligned almost vertically along the direction of gravity (Fig. 2a, b). Hence, load transmission through the contact lines would result in the local stress minimum beneath the apex of narrow source deposited piles. The existence of such preferred directions of propagation explains why some phenomenological yet insightful continuum models[35-38,41] can predict the stress dip in granular piles. Interestingly, in piles deposited from a narrow source but with monodisperse particles, the contact lines are remarkably different compared to the polydisperse case (Fig. 2c) and vertical orientation of the contact lines in the core regions explains the absence of central stress minimum reported in previous studies[40,45,46]. In silos with frictional walls, the contact lines show a clear curvature towards the lateral walls (Fig. 3a, c). And, in silos made up of smooth frictionless walls but with frictional particles where the Janssen effect is absent, the contact lines are almost vertically oriented in the direction of gravity (Fig. 3b). Hence, in silos with frictional walls, load transmission along the contact lines would result in the vertical load screening effect with depth i.e. Janssen effect[7,27,28]. Later, using a simple model of force transmission, we corroborate the claims we have made above, and directly relate the contact lines to the stress anomalies observed in pile and silo packings. We also find that, the contact line patterns are independent of the type of contact model used, various particle-particle interaction parameters, and deposition procedure (see Supplementary Note 2). Hence, the contact line representation is a useful qualitative measure to understand the spatial structure in granular media. Next, we discuss the emergence of these spatial patterns in detail by studying the time evolution of pile and silo packings.

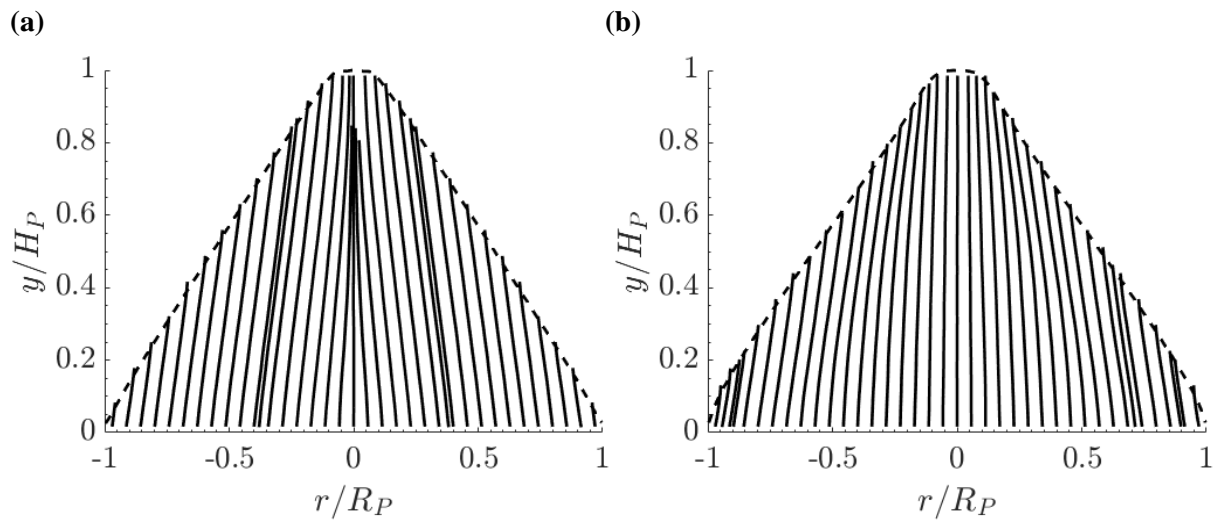

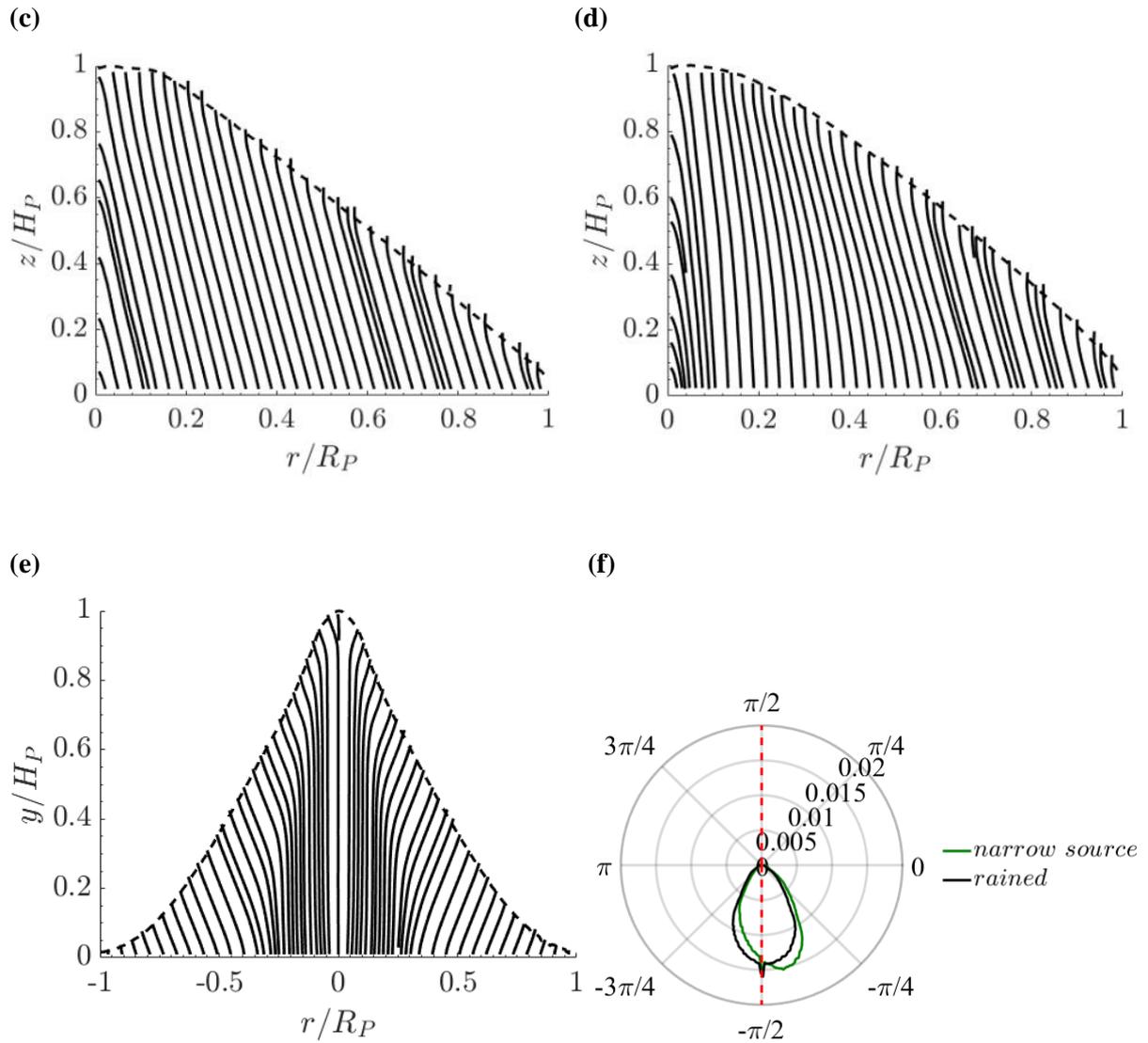

**Figure 2: Spatial structure of granular piles.** Contact lines of 2d and 3d piles. **(a)** 2d piles, deposited from a narrow source. **(b)** 2d piles, deposited by raining. **(c)** 3d piles, deposited from a narrow source. **(d)** 3d piles, deposited by raining. **(e)** 2d pile of monodisperse particles deposited from a narrow source. In all cases, dashed lines represent the average height of the free surface. **(f)** Orientation distribution of $\vec{R}$ (Fig. 1) in 2d piles, here we have considered only the core region of the right-hand side of the pile. The core region is defined as particles within $r = 0$ and $r = 0.3R$, where, $R$ is the half width of the pile. The details of the system and number of configurations studied are given in Supplementary Note 2.

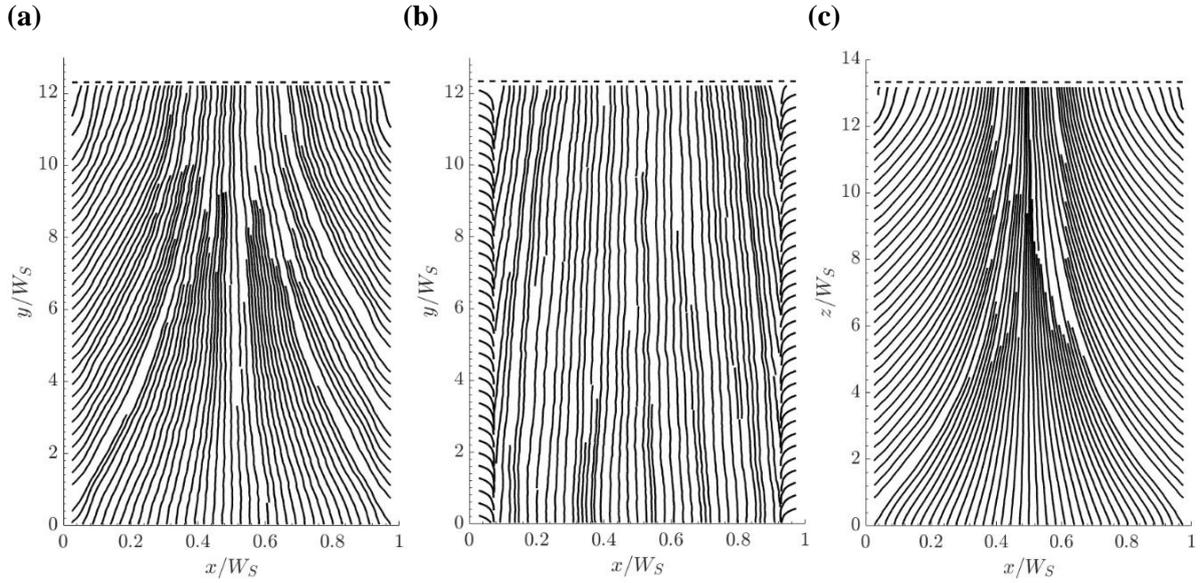

**Figure 3: Spatial structure of granular silos.** Contact lines of 2d and 3d silos. **(a)** Contact lines of 2d silos with rough frictional walls. **(b)** Contact lines of 2d silos with smooth frictionless walls, but the particles are frictional. **(c)** Contact lines of 3d silos with rough frictional walls. In all cases, dashed lines represent the average height of the free surface, and the width of the silos, $W_s = 20 d_p$ and the average height of the free surface, $H_s \approx 12 W_s$. The details of the system and number of configurations studied are given in Supplementary Note 2.

In piles, the streamlines of the displacement field of particles reveal interesting patterns of evolution depending on the method of deposition (see Supplementary Video 6-8) (details of Supplementary Videos are given in Supplementary Note 4). In narrow source deposited piles, we find a strong horizontal outward flow from the centre (see Supplementary Video 6) and in rained piles, a perceptible flow is observed only along the surface (see Supplementary Video 7). We also studied the evolution of piles by sequentially labelling the particles based on their time of deposition (see Supplementary Video 1,2), in narrow source deposited piles, the spread of newly deposited layers of particles in the horizontal direction due to the outward flow can be clearly seen. In addition, the time evolution of the contact lines also correlates well with the layering patterns induced by the nearly horizontal flow (see Supplementary Video 1). Hence, the strong horizontal component of the particle flow at the apex of the pile results in noticeable anisotropy in the core region of the pile (Fig. 2d), which is preserved in subsequent layers of newly deposited particles. Such changes in microstructure caused by externally imposed shear or deformation in granular media is well known[3,5,6,7,42]. Also, recent continuum modelling and discrete simulation studies find strong anisotropy in the core region of the pile as suggested by our results[25,43] (Fig. 2d). We note that, the above conclusion is in complete contrast to the surface avalanches hypothesis suggested by many previous studies[21-25,35-38,41,42]. Curiously, we find that, the emergence of structural pattern in monodisperse piles is primarily driven by surface avalanches (see Supplementary Videos 3,8) and is remarkably different compared to the polydisperse case. Though local ordering or crystallization effects are important in monodisperse packings[47,48], here we have shown that, surface avalanches in granular piles can lead to a maximum of stress beneath the apex.

The emergence of the spatial order in silos is also interesting, we explain this by studying the evolution of the displacement field of the particles and the contact lines (see Supplementary Videos 4,5,9,10). In silos with frictional walls, the spatial order emerges from particle rearrangements near the growing free surface. We find clear particle movements towards the lateral walls in the growing free surface region of the packing (see Supplementary Video 9). The microstructural rearrangements caused by the surface flow is reflected in the evolution of contact lines with time (see Supplementary Video 4). However, in silos made of smooth frictionless walls but with frictional particles, in addition to particle movements towards the lateral walls in the growing free surface region, we find slow and continuous nearly vertical flow along the entire height of the column (see Supplementary Video 10). The time evolution of particle displacements and contact lines clearly suggests that, the slow continuous vertical flow determines the microstructure of the packing in the final static state (see Supplementary Video 5,10). Hence, we conclude that, frictional resistance of the lateral walls hinders the slow vertical flow, and particles movements towards the lateral walls in the growing free surface region determine the spatial structure. We next show how the observed structural patterns can help us understand central stress minimum and Janssen effect, using a simple model of force transmission. It is important to note that, here we do not attempt to develop an exact model of force transmission, our aim is only to show the effect of spatial structure on the mechanical response of granular materials.

We use the simplest possible model of force transmission in granular materials i.e. the weight of the particle is assumed to be transmitted to the supporting base by a random walk. As the information of spatial structure is unknown, we use the information of the contact network from DEM generated packings in the random walk studies. Our objective here is to show that, only the information of the contact network (or the spatial structure of it) can be used to explain the central stress minimum and Janssen effect in granular media, using a force transmission model as simple as a random walk. Here, the random walk can propagate only in the direction of gravity; we refer to this as the Random Walk (RW) model. In RW, the weight of a bulk particle is transmitted to the supporting base (boundary particles) through random paths (see Methods section for details), and the fraction of the total weight of the pile reaching a boundary particle is found and is averaged over many independent realizations. Though the random walk is an oversimplified model of force transmission, remarkably it predicts the central stress minimum in piles reasonably well for polydisperse piles (Fig. 4a-d). Interestingly, we find that the RW predictions in piles made up of particles with large spring stiffness values or Hertz spring are in very close agreement with DEM results (see Supplementary Figures 7,8), which further suggests that the RW is a reasonable model of force transmission in realistic granular piles. However, in monodisperse piles, though the RW explains the presence of a maximum of stress beneath the apex similar to DEM estimates, a close agreement, as observed in polydisperse piles, is lacking (Fig. 4h), the reason is unclear to us, but can be caused by local ordering or crystallization effects[47,48] as discussed before. We base our above claim on point force experiments on 2d disk packings[4,5], where force transmission in ordered packings was shown to be dramatically different compared to disordered packings. Next, to the show the effect of spatial structure on RW prediction in silos, we compare packings with frictional and frictionless walls. As shown in Fig. 4e,g, the RW in the frictional wall-bounded silo packing shows strong signatures of Janssen saturation compared to packings with smooth frictionless walls (Fig. 4f), this can be better understood by contrasting the spatial structures of the packings (Fig. 3a,b). We note that, load transfer due to frictional interactions are important in wall-bounded packings like silo[17,27,49], and RW does not account for friction between particles, hence we do not observe a close agreement between RW and DEM predictions in the silo with frictional walls compared to the frictionless case. The effect of wall roughness and width of silo are discussed in Supplementary Note 2.

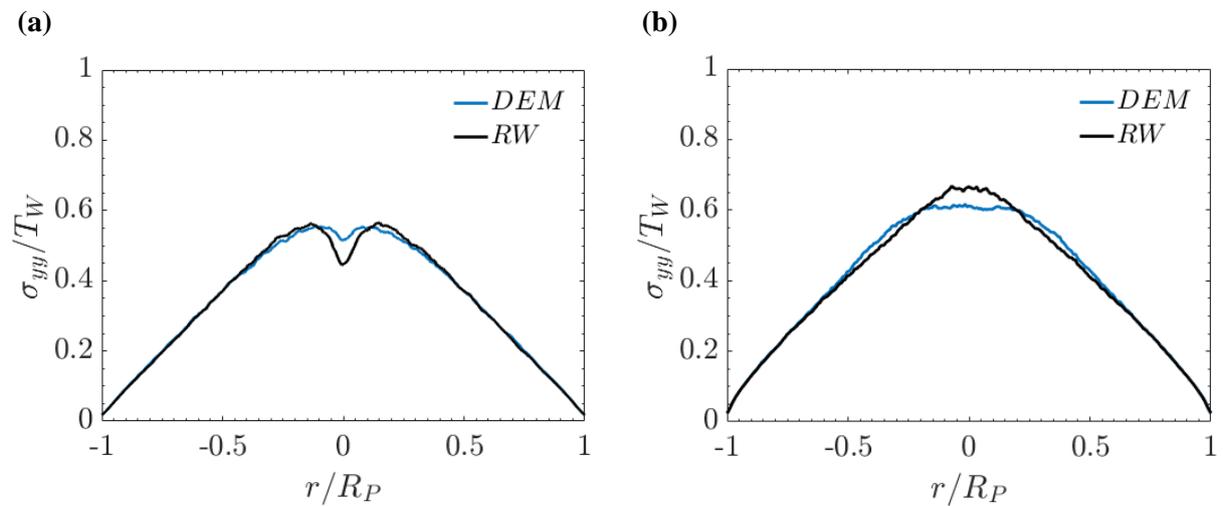

(a)  (b)

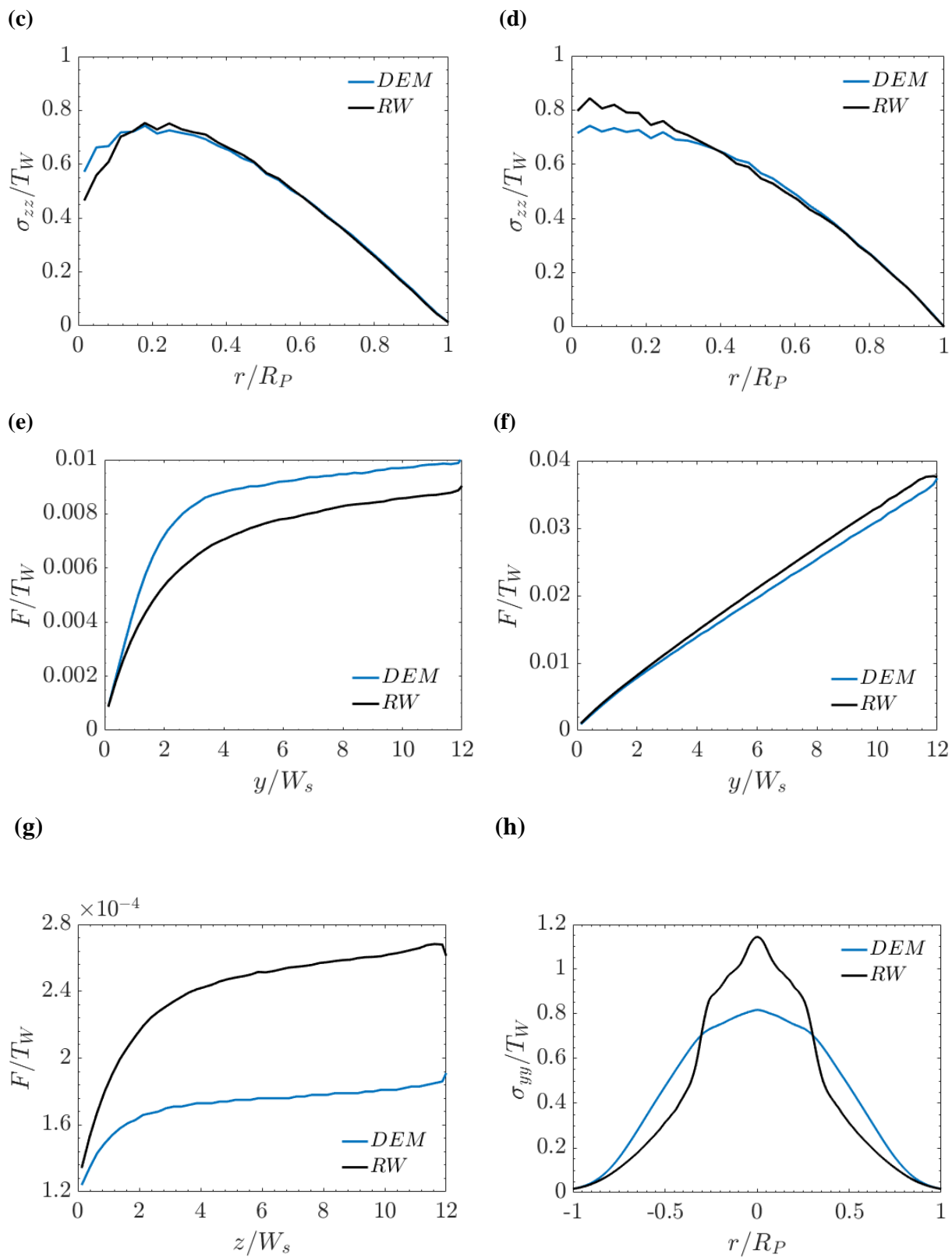

**Figure 4: The random walk model.** The predications of RW compared to the estimates from particle dynamics simulations. Here, σ is the normal stress measured at the supporting base of the pile and $F$ is the average contact force at a given depth from the free surface of the silo ($F_n$ in DEM, and $F^{RW}$ in RW (see Methods section for details)), $T_W$ is the total weight of the packing. **(a)** 2d piles deposited from a narrow source. **(b)** 2d piles deposited

by raining. **(c)** 3d piles deposited from a narrow source. **(d)** 3d piles deposited by raining. **(e,f)** 2d silos with rough frictional walls **(e)** and smooth frictionless walls **(f)**. **(g)** 3d silos with rough frictional walls. **(h)** 2d piles made up of monodisperse particles deposited from a narrow source. The averaging procedure and number of configurations studied are given in Supplementary Note 2.

To clearly show how a random walker's movement is determined by the spatial structure of the packing? We now compare the average trajectories of random walks with contact lines. In RW, the total weight transmitted through a contact ($F^{RW}$) can be found (see Methods section for details). Hence, each contact can be weighted by $F^{RW}$, and we define the vector field based on the weighted contact vectors. We then construct streamlines based on weighted contact vector field and refer to this as random walk lines (see Supplementary Note 4). As expected, the random walk lines agree almost exactly with contact lines (see Supplementary Fig. 11). Interestingly, in granular piles made up of polydisperse particles, we also find that, streamlines based on contact vectors weighted by the normal contact force ($F_n$), are also in close agreement with the contact lines (see Supplementary Fig. 12a-d), and we refer to this as force lines. It is important to note that, in the random walk model, the direction of movement is independent of the direction of the previous step (spatially uncorrelated). Hence, long-ranged structural correlations or force chains or arches are mechanically not significant in the emergence of central stress minimum in granular piles, unlike previously thought[21-24,35-38]. However, in monodisperse piles, as suggested by the RW predictions, we find that the force lines and contact or random walk lines are similar only the core region, and show drastic difference away from the centre (see Supplementary Fig. 12e). In silos too, we find that the contact lines and random walk lines agree almost exactly (see Supplementary Fig. 11f-h), and as explained before, the force weighted contact lines show stronger curvature compared to the contact lines due to frictional interactions of particles with lateral walls (see Supplementary Fig. 12f-h). Hence, using only the information of the contact network and with random walk as a model of force transmission, we have clearly shown the significance spatial variations of local microstructure on some curious cases of mechanical behaviour in granular materials. Finally, we discuss how the information of spatial structure can be used in modelling force transmission in granular piles.

As explained previously, in the case of granular piles, the RW is a useful model of force transmission. Hence, we explain how the spatial structure can be represented in the continuum analogue of the RW. In the continuum limit, the RW resembles the drift-diffusion equation[50,51] (see Methods). The transmission of the weight of a grain or applied load ($W$) is given by,

$$\frac{\partial W}{\partial t} = -\nabla \cdot (\mathbf{v}W) + \nabla \cdot (\nabla(\mathbf{D}W))  \qquad (1)$$

Where, time ($t$) represents the spatial coordinate ($y$ in 2d and $z$ in 3d) in the direction of gravity, $\vec{g}$. Here, the information of the spatial structure is represented by spatially varying, drift rate, $\mathbf{v}(x, y)$ and diffusion constant, $\mathbf{D}(x, y)$. The functions $\mathbf{v}(x, y)$ and $\mathbf{D}(x, y)$ are difficult to find, only in the simplest of cases, as in a rained pile of grains (Fig. 2b,d) or a silo packing with frictionless walls (Fig. 3c) can be modelled as spatially homogenous.

**Conclusion**

By visually presenting the spatial variation of local microstructure of the complete packing, we have shown the existence of large-scale spatial structure in granular media. Based on the spatial structure of pile and silo packings, we have provided fundamental explanations to some long standing and puzzling mechanical behaviours in granular media. Moreover, we have shown that, the spatial structure can vary dramatically depending on the method of deposition, boundary conditions and size distribution of particles, and consequently the mechanical behaviour too. The mechanical significance of spatial structure in other disordered media like foams, emulsions and even biological populations are worth investigating.

**Acknowledgements**

I sincerely thank Professor Prabhu R Nott for insightful discussions and useful critiques. I am profoundly grateful to N.Marayee, P.Rajamani, K.P.Dhanasekaran, A.K.K.Arvind and Mani for their encouragement and support. I also thank R.S.Veeraraahavan and S.Karthick for their hospitality.

**Methods**

**1. Random walk model**

In the random walk model, the weight of individual particles is transmitted to the boundaries through contacts by a random walk. Here, we only choose contact directions with a positive vertical component in the direction of gravity. From every particle, numerous paths (through the contact network) are possible for a random walk to reach the supporting base. And, enumerating all possible paths reaching the supporting base is computationally prohibitive. Hence, in this study, we use a statistical approach, we randomly sample a subset of all possible paths with $RW^i$ random walks from the given bulk particle, $i$. The fraction of the weight of a bulk particle $i$ transmitted to a boundary particle $j$, $F_{ij}$ is given by,

$$F_{ij} = \frac{RW_j^i \times W_i}{RW^i} \qquad (2)$$

Where, $W_i$ is the weight of a bulk particle $i$ and $RW_j^i$ is the number of random walks reaching a boundary particle $j$ starting from a bulk particle $i$.

Hence, in a pile made up of $N$ particles, the fraction of the total weight of the pile supported by a boundary particle $j$, $F_j^{RW}$ is given by,

$$F_j^{RW} = \sum_{i=1}^{N} \frac{RW_j^i \times W_i}{RW^i} \qquad (3)$$

We find that, $RW^i = 25$ is enough to achieve reasonable statistics, and the estimates of stress does not change with increasing values of $RW^i$ (see Fig. 5).

Similarly, the total weight transmitted through a contact $c$ in the random walk model $F_c^{RW}$ is given by,

$$F_c^{RW} = \sum_{i=1}^{N} \frac{RW_c^i \times W_i}{RW^i} \qquad (4)$$

Here, $RW_c^i$ is the number of random walks passing through a contact $c$ starting from a bulk particle $i$.

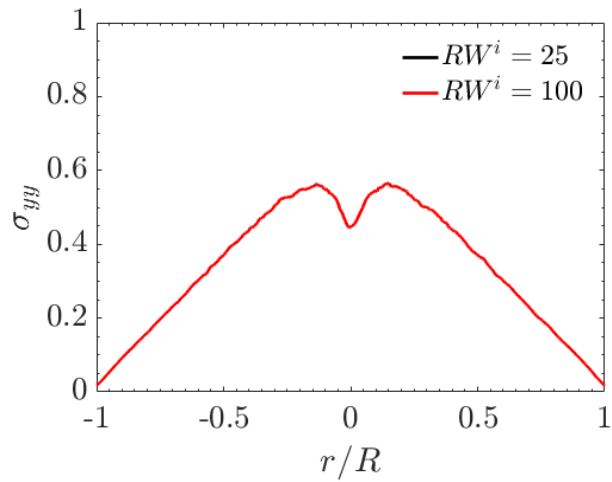

**Figure 5:** The effect of number of paths ($RW^i$) sampled randomly for a bulk particle. We find that, for every bulk particle $i$, $RW^i = 25$ is a sufficient, and increase in the number of paths sampled does affect the RW predictions.

## 2. Continuum approximation of the random walk model in disordered media

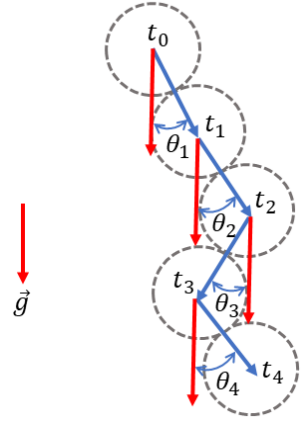

**Figure 6:** Schematic of a random walker moving through a disordered packing in the direction of gravity. Here, $\vec{g}$ is the unit vector in the direction of gravity.

We model force transmission in granular materials as a random walker moving through a disordered lattice. At each time step $\tau$, the random walker can move a distance $\delta$ ($\approx d_p$) in one of the $N$ possible directions given by $\theta_i$, where $\theta_i \in \{\theta_1, \theta_2, \theta_3 \dots \theta_N\}$ and $\{0 \leq \theta_i < 2\pi\}$[50]. The probability of transition, which depends on the current position $(x, y)$ is given by $P_i(x, y)$, and $\sum_{i=1}^{N} P_i(x, y) = 1$. Here, we consider forward progression in time ($t$) as movement in space along the direction of gravity, $y$ in 2d and $z$ in 3d. The weight of a grain ($W$) at position $(x, y)$ after a forward time step is given by[50],

$$W(x, y, t) = \sum_{i=1}^{N} \Big(W(x - \delta \sin(\theta_i), y - \delta \cos(\theta_i), t - \tau) p_i(x - \delta \sin(\theta_i), y - \delta \cos(\theta_i))\Big)$$

$$+ W(x, y, t - \tau)\left(1 - \sum_{i=1}^{N} p_i(x, y)\right) \quad (4)$$

As shown in chapter 2 of reference 50, in the continuum limit i.e., $\delta$ and $\tau \to 0$, the above equation can be written as,

$$\frac{\partial W}{\partial t} = -\nabla \cdot (\mathbf{v}W) + \nabla \cdot \left(\nabla(\mathbf{D}W)\right) \tag{5}$$

Where, $\mathbf{v}(x,y)$ and $\mathbf{D}(x,y)$ are spatially varying drift rate and diffusion constant respectively. And are given as,

$$\mathbf{v} = \begin{pmatrix} v_1 \\ v_2 \end{pmatrix} \quad and \quad \mathbf{D} = \begin{pmatrix} d_{11} & d_{12} \\ d_{12} & d_{22} \end{pmatrix} \tag{6}$$

Equation (5) is reminiscent of the two-dimensional drift-diffusion equation[50,51]. Here, the drift rate and diffusion coefficient depend on the spatial position, typically unknown functions, and are described by the contact lines in Fig. 2. Hence, given the information of local microstructure of the packing, $\mathbf{v}(x,y)$ and $\mathbf{D}(x,y)$, the base stress distribution in gravity deposited granular piles can be predicted using equation (5).

# Supplementary Material

**Supplementary Note 1**

## 1. Particle dynamics simulations

The Discrete Element Method (DEM) is a particle dynamics simulator with an elastoplastic interaction force, which is widely used for computational simulation of granular statics and flow (Supplementary Refs[1]). Our simulations were conducted using the open source molecular dynamics package LAMMPS (Supplementary Refs[2]), and the contact model and its DEM implementation are described in Supplementary Refs[3]. In DEM the particles are treated as deformable, and their interaction forces are calculated from the normal overlap and tangential displacement post contact. The dissipative interaction is modelled by spring-dashpot modules for the normal and tangential directions (Supplementary Fig. 1), and an additional Coulomb slider in the latter to incorporate a rate-independent frictional force, an important feature of granular materials. For a pair of spheres $i, j$ of radii $R_i, R_j$ at positions $x_i, x_j$ in contact, the overlap is

$$\delta \equiv R_i + R_j - |x_{ij}| \tag{1}$$

where $x_{ij} \equiv x_i - x_j$; the particles are in contact only when the overlap is positive. The components of the relative velocity normal and tangential to the point of contact are

$$\mathbf{v}_{n_{ij}} = (\mathbf{v}_{ij} \cdot \mathbf{n}_{ij}) \mathbf{n}_{ij} \tag{2}$$

$$\mathbf{v}_{t_{ij}} = \mathbf{v}_{ij} - \mathbf{v}_{n_{ij}} - (\boldsymbol{\omega}_i R_i + \boldsymbol{\omega}_j R_j) \times \mathbf{n}_{ij} \tag{3}$$

where $\mathbf{n}_{ij} \equiv x_{ij}/|x_{ij}|$ is the unit normal from $j$ to $i$, $\mathbf{v}_{ij} \equiv \mathbf{v}_i - \mathbf{v}_j$, and $\boldsymbol{\omega}_i, \boldsymbol{\omega}_j$ are the rotational velocities of particles $i$ and $j$. The tangential spring displacement $\boldsymbol{u}_{t_{ij}}$ is initiated at the time of contact and can be calculated by integrating,

$$\frac{d\boldsymbol{u}_{t_{ij}}}{dt} = \mathbf{v}_{t_{ij}} - \frac{(\boldsymbol{u}_{t_{ij}} \cdot \mathbf{v}_{ij}) x_{ij}}{|r_{ij}^2|} \tag{4}$$

The second term represents rigid body rotation around the point of contact and ensures that $\boldsymbol{u}_{t_{ij}}$ lies in the tangent plane of contact.

For simplicity, the springs are assumed to be linear (Hookean). Previous studies (Supplementary Refs[2]) have shown that employing non-linear springs that corresponds to Hertzian contact makes no qualitative difference. We also find that the use of a non-linear spring leads to similar results (see Supplementary Note 2). The normal and tangential forces imparted on $i$ by $j$ are

$$\boldsymbol{F}_{n_{ij}} = k_n \, \delta_{ij} \, \boldsymbol{n}_{ij} - \gamma_n \, m_{\text{eff}} \, \boldsymbol{v}_{n_{ij}} \tag{5}$$

$$\boldsymbol{F}_{t_{ij}} = \begin{cases} -k_t \, \boldsymbol{u}_{t_{ij}} - \gamma_t \, m_{\text{eff}} \, \boldsymbol{v}_{t_{ij}} & \text{if } |\boldsymbol{F}_{t_{ij}}| < \mu \, |\boldsymbol{F}_{n_{ij}}| \\ -\mu \, |\boldsymbol{F}_{n_{ij}}| \, \boldsymbol{v}_{t_{ij}} / |\boldsymbol{v}_{t_{ij}}| & \text{otherwise} \end{cases} \tag{6}$$

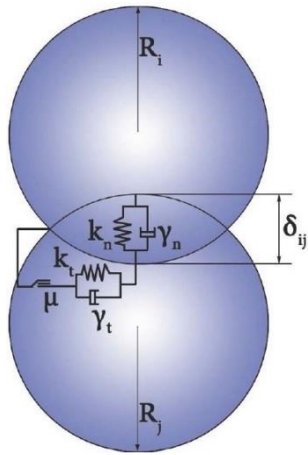

| Parameter | Value |
|---|---|
| $k_n$ | $10^5$ $(N/m)$ |
| $k_t$ | $\frac{2}{7} k_n$ |
| $\gamma_n$ | $2750 \, (1/sec)$ |
| $\gamma_t$ | $\frac{1}{2} \gamma_n$ |

**Supplementary Figure 1:** Schematic of the soft particle interaction model between spheres of radii $R_i$ and $R_j$. The values of parameters used in the model are given in the table, where $\vec{g}$ is the gravitational acceleration on earth, and $m_p$ is the mass of a particle of diameter $d_p$.

where $k_n$ and $k_t$ are the normal and tangential spring stiffness coefficients, $\gamma_n$ and $\gamma_t$ the corresponding damping coefficients, $\mu$ is the coefficient of friction for the Coulomb slider, and $m_{\text{eff}} \equiv m_i \, m_j / (m_i + m_j)$ is the effective mass of the two spheres. The velocities and positions of the particles are updated by integrating Newton's second law,

$$m_i \dot{\boldsymbol{v}}_i = \sum_j \boldsymbol{F}_{ij} + \boldsymbol{F}_i^{\text{ext}}, \quad I_i \dot{\boldsymbol{\omega}}_i = -\frac{\sum_j x_{ij} \times F_{ij}}{2} \tag{7}$$

where pairwise additivity of the interaction forces is assumed, and $\boldsymbol{F}_i^{\text{ext}}$ is the external force (such as gravity).

For the linear spring-dashpot-slider model, the time of contact is (Supplementary Refs[3])

$$t_{\text{coll}} = \pi (2k_n/m - \gamma_n^2/4)^{-1/2}. \tag{8}$$

The choice of the normal spring stiffness coefficient determines the collision time between two particles. The simulation time step is chosen such that each collision is resolved accurately, and the choice of $\Delta t = t_{\text{coll}}/50$ is found to be sufficiently small[6] (Supplementary Refs[3]). Since the collision time decreases with increasing spring stiffness $k_n$, it is standard practice to optimize the value of $k_n$ such that it is large enough for the macroscopic behaviour to mimic that of hard particles, and the time step is large enough for the computations to be tractable. The parameters used in the simulations are listed in Supplementary Fig. 1.

The values of $k_n$, $k_t$ and $\gamma_t$ was chosen based on previous studies[6] (Supplementary Refs[3]) that have attempted to model hard grains such as glass beads and sand. The value of $\gamma_n$ chosen is such that the normal coefficient of restitution is 0.7. In all our computations, $\mu$ is set to 0.5. The 2d simulations were conducted by placing spheres in a plane and allowing movement only within the plane.

In our simulations, the mean particle size is 1cm $(d_p)$, and the particle sizes were chosen from a uniform distribution with lower and upper limits of $0.8d_p$ and $1.2d_p$ respectively. The walls were constructed with particles of diameter $d_p$ set in a close packed linear (2d) or triangular (3d) lattice. In all the simulations, the constants characterizing grain-wall interactions are the same as those for grain-grain interactions.

**Supplementary Note 2**

**1. Method of deposition**

**1.1 Piles**

**1.1.1 Narrow source**

**(a) Deposition from a vertical narrow region**

In the narrow source deposition procedure, non-overlapping particles were randomly created in a narrow region (coloured grey in Supplementary Fig. 4), and were allowed to fall under the influence of gravity, $\vec{g}$. Here, the region of deposition is randomly filled with non-overlapping particles to a volume fraction of 0.1 (area fraction of 0.2 in 2d), and the next set of particles are created after the current set has moved out of the deposition region under the influence of gravity. In 2d, the length and width of the deposition area is, $L_D = 50d_p$, and, $W_D = 10d_p$ respectively (see Supplementary Fig. 4a). In 3d, the deposition region was a cylinder with length, $L_D = 60d_p$ and radius, $R_D = 7d_p$. The deposition area was located at a fixed height, $H_D = 60d_p$ (in 2d) ($50d_p$ in 3d) from the supporting base of the pile. The deposition was continued till a fixed number of particles were poured and the simulations were continued till the kinetic energy per particle decreased to a value of $10^{-12}J$. The narrow source deposited piles described in Fig. 2 were created by this procedure. The dimensions of the piles created are given in Supplementary Table 1. Unless specified otherwise, narrow source deposited piles described in this study were created by this procedure.

| Method of deposition | Height, $H_P$ | Radius, $R_P$ | Angle of repose, $\phi_r$ |
|---|---|---|---|
| 2d, Narrow source | 42 | 144.5 | 16.20 |
| 2d, Narrow source (1.1.1b) | 45 | 143 | 17.47 |
| 2d, Hopper | 41 | 148 | 15.48 |
| 2d, Rained | 42 | 125 | 18.57 |
| 3d, Narrow source | 34 | 94 | 19.89 |
| 3d, Rained | 28 | 90 | 17.28 |

**Supplementary Table 1:** Dimensions of the piles created by different methods of deposition. In all cases, the radius of the pile $(R_p)$ is the minimum distance from the centre at which the average height of the pile is $\approx 1d_p \pm (d_p/4)$.

**(b) Deposition from a vertical narrow region with reduced length and lesser height of deposition**

The effect of height of the deposition region on the central stress minimum in 2d piles deposited from a narrow source was studied. To reduce the effects of impact force generated by the newly poured particles, we deposited from a region with $H_D = 45d_p$, and $L_d \times W_d (10d_p \times 10d_p)$ (see Supplementary Fig. 4a), and the fraction of area filled with particles is 0.4. We note that, the height of the pile $(H_p = 42d_p)$ is comparable to the $H_D$. In addition, the reduced length of the deposition region (by a factor of 5), considerably reduces the impact force generated due to the newly poured particles. As given in Supplementary Table 1, the reduced effect of impact force results in a slightly steeper pile. We find that, the spatial structure patterns and stress profiles measured at the supporting base are similar to the previous case (see Supplementary Fig. 2). However, in piles made of stiffer particles $(k_n = 10^6)$, increasing heights of deposition affects the shape of the free surface of the pile. Hence, to construct piles with stiffer particles, we used a lower height of deposition $(H_D = 45d_p)$. The dimensions of the piles created are given in Supplementary Table 1.

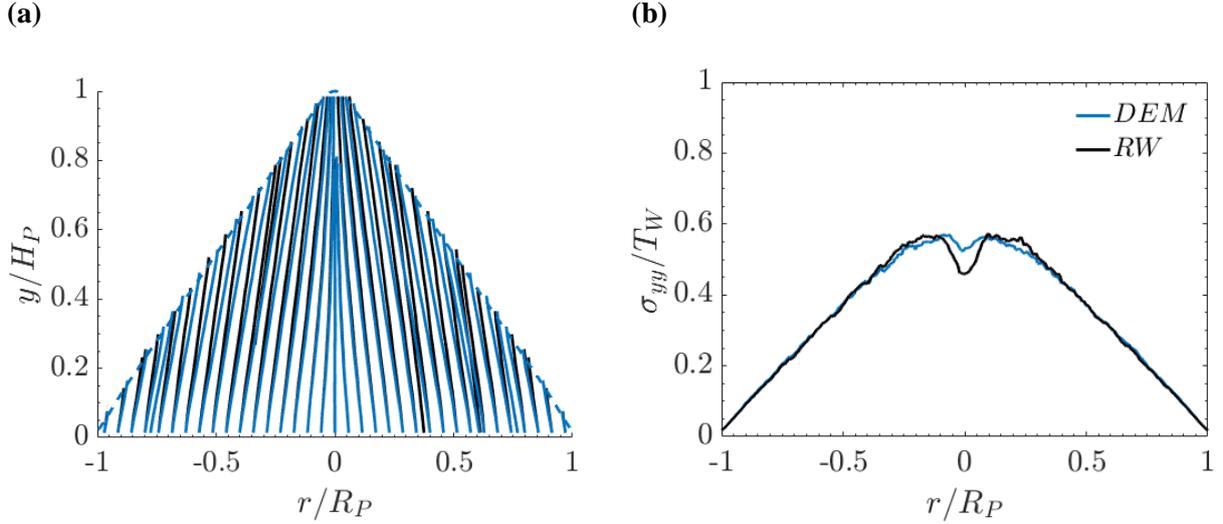

**Supplementary Figure 2:** Spatial structure of piles created by depositing particles from a vertical narrow region with reduced length and lesser height of deposition. Dashed lines represent the average height of the free surface of the pile. **(a)** Contact lines (black) and force lines (blue). **(b)** RW predications compared with the estimates from particle dynamics simulations. $\sigma_{yy}$ is the normal stress measured at the supporting base and $T_W$ is the total weight of the pile.

### (c) Deposition from hopper

Experimental studies on central stress minimum have created narrow source deposited piles by pouring particles from a hopper[16-22,24]. Hence, we also created 2d piles by pouring particles from a hopper geometry with a funnel-shaped exit as shown in Supplementary Fig. 4c. The hopper was filled by raining the particles, and the kinetic energy of the packing was drained till it reached a nearly static state. Once the kinetic energy per particle decreased to a value of $10^{-12}J$ the particles were allowed to move out of the hopper exit under the influence of gravity. We find that the contact lines and the stress profiles measured at the base are similar to piles deposited from a narrow region discussed previously (see Supplementary Fig. 3). The dimensions of the hopper shown in Supplementary Fig. 2c are $L_1^H = 80d_p, L_2^H = 145d_p, L_3^H =$

$50d_p$, $W_d = 300d_p$, $R_o = 10cm$ and $H_d = 50d_p$. The dimensions of the piles created are given in Supplementary Table 1.

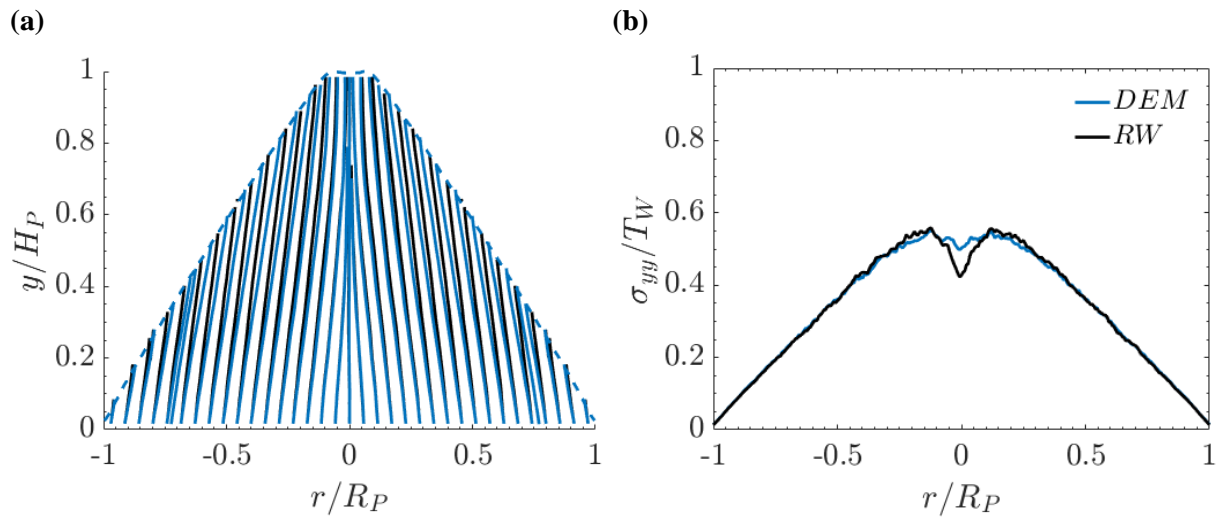

**Supplementary Figure 3:** Spatial structure of piles created by depositing particles from a hopper. **(a)** Contact lines (black) and force lines (blue). Dashed lines represent the average height of the free surface of the pile. **(b)** RW predications compared with the estimates from particle dynamics simulations. $\sigma_{yy}$ is the normal stress measured at the supporting base and $T_W$ is the total weight of the pile.

### 1.1.2 Rained

In the rained deposition method, the width of the deposition ($W_D$) region was equal to the radius of the heap, $R_P$ (see Supplementary Fig. 4b). Other aspects of the deposition procedure are similar to the narrow source deposition method. In 3d, the deposition region is a cylinder of length $L_D$, and radius $R_D$, with dimensions $10d_p \times 90d_p$ ($L_D \times R_D$). The rained piles described in Fig. 2 were created by this procedure. The dimensions of the piles created are given in Supplementary Table 1. Unless specified otherwise, rained piles described in this study were created by this procedure.

**(a)**

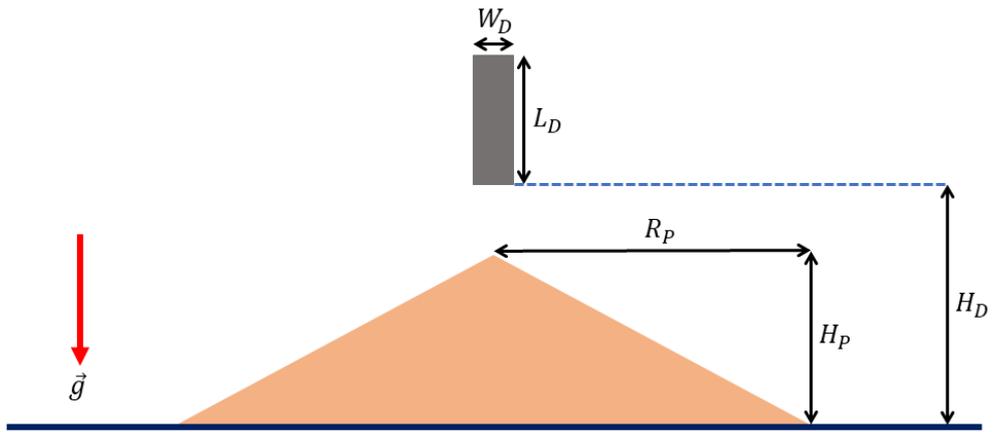

**(b)**

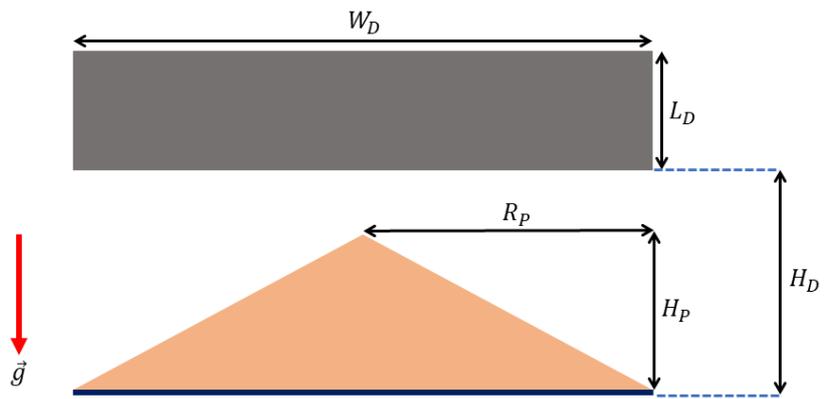

**(c)**

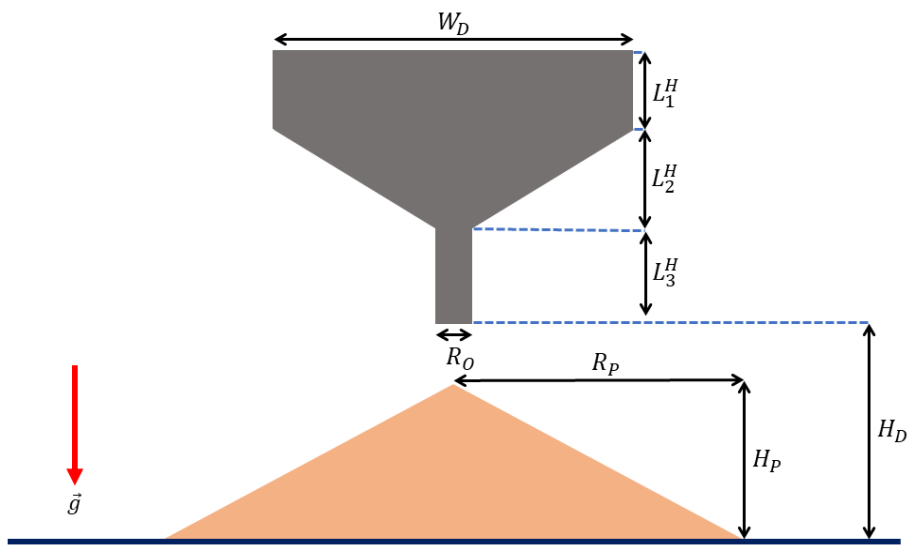

**Supplementary Figure 4:** Schematic of different methods of deposition. **(a)** Narrow source deposition **(b)** deposition by raining **(c)** deposition from a hopper. In all cases, the region of deposition in shown in grey.

## 1.2 Silo

In silos, the particles were rained from the top of the silo, as shown in Supplementary Fig. 5. The length and width of the deposition region in 2d is $100d_p$ ($L_D$) and $18d_p$ ($W_D$) respectively, in 3d the deposition region was a rectangular volume of length, $L$, width $W$ and depth $D$, with dimensions $100d_p \times 18d_p \times 23d_p$ ($L \times W \times D$). In silo too, the region of deposition is randomly filled with non-overlapping particles to a volume fraction of 0.1 (area fraction of 0.2 in 2d), and the next set of particles are created after the current set has moved out of the deposition region under the influence of gravity. As stated before, in rough walled silos, the walls were constructed with particles of diameter $d_p$ set in a close packed linear (2d) or triangular (3d) lattice. In case of smooth walls, the walls are flat featureless planes. As shown in Supplementary Figure 5b,c, in case of frictional walls the contact lines and RW predictions are similar for both rough and smooth walled silos. The dimensions of the silo packings created are given in Supplementary Table 2.

| Type of silo | Fill height, $H_S$ | Width, $W_S$ |
|---|---|---|
| 2d, Rough frictional walls | 246 | 20 |
| 2d, Smooth frictional walls | 247.5 | 20 |
| 2d, Smooth frictionless walls | 246.5 | 20 |
| 2d, Rough frictional walls (Wide) | 490.0 | 40 |
| 3d, Rough frictional walls | 266.5 | 20 |

**Supplementary Table 2:** Dimensions of the silo packings.

**(a)**

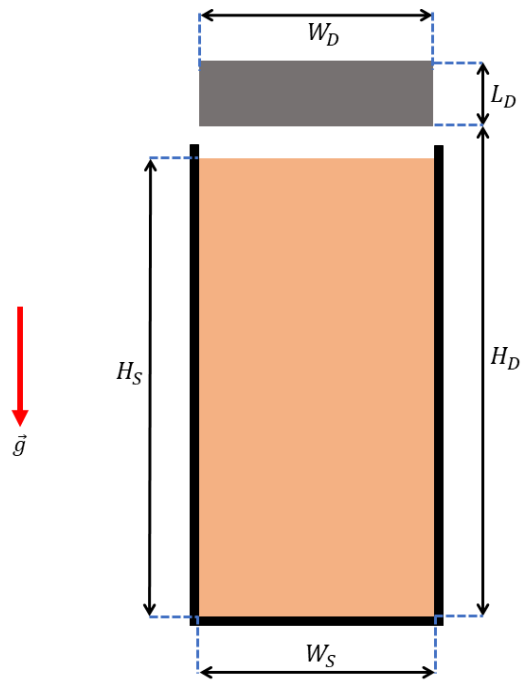

**(b)**

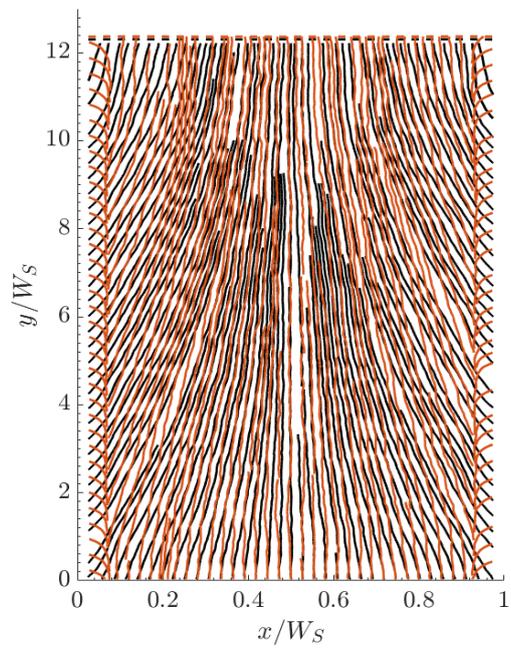

**(c)**

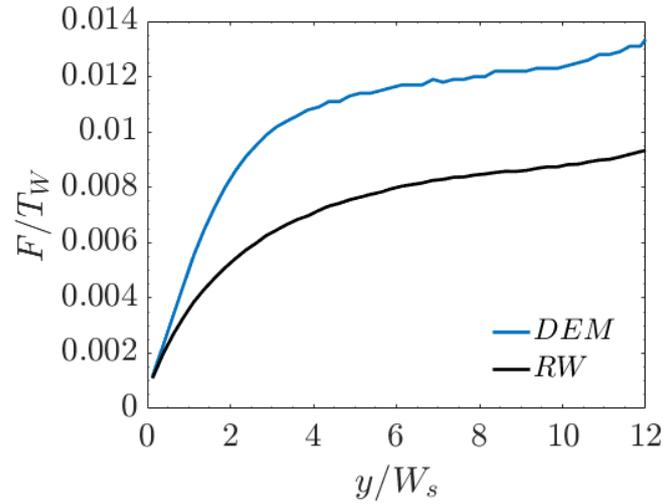

**Supplementary Figure 5:** Schematic and contact lines of a silos **(a)** Schematic of a silo packing, with the region of deposition shown in grey. **(b)** Comparison of contact lines in 2d silos with rough frictional walls (black) and smooth frictional walls (brown). **(c)** Predications of RW compared to the estimate from particle dynamics simulations in 2d silos with smooth frictional walls. $F$ is the average contact force at a given depth from the free surface of the silo ($F_n$ in DEM, and $F^{RW}$ in RW (see Methods section for details)), $T_W$ is the total weight of the system.

### 1.2.1 Width of the silo

To verify the predictions of RW and the profile of contact lines in wider silo packings, we studied 2d silo packings of width, $W_s = 40d_p$. The contact line profiles are similar to 2d silos of width, $W_s = 20d_p$ (Supplementary Fig. 6a), and the predictions of RW are also comparable with estimates from particle dynamics simulations (Supplementary Fig. 6b).

**(a)**                                                    **(b)**

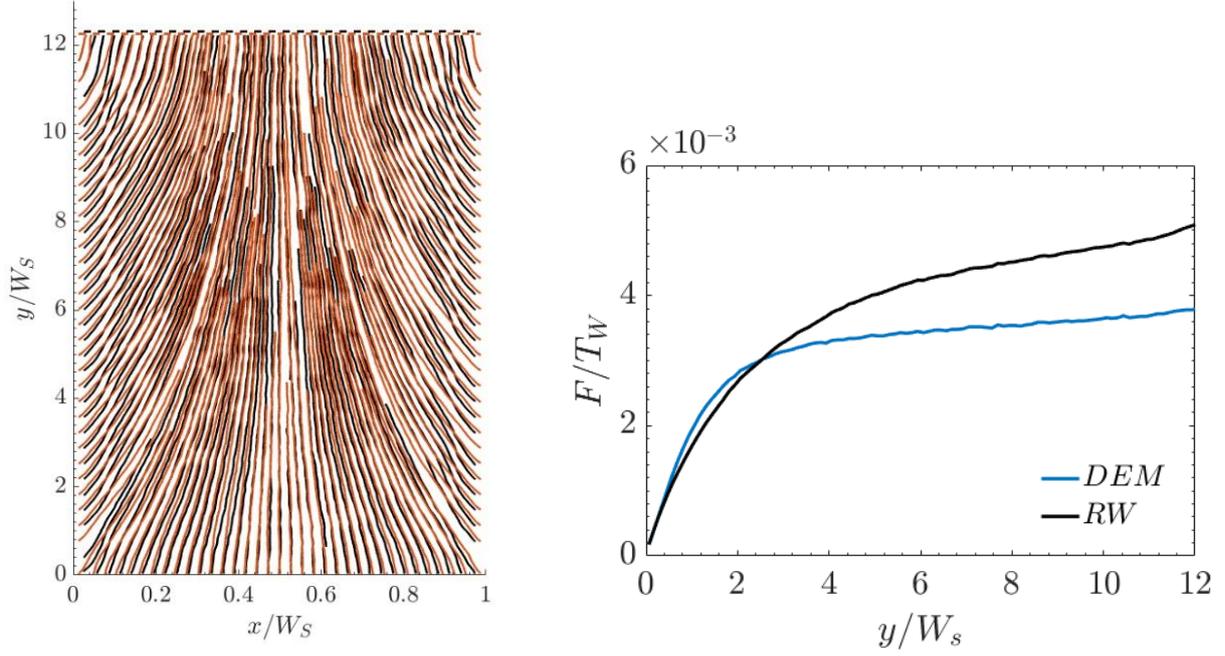

**Supplementary Figure 6:** Spatial structure of 2d silos with different widths ($W_s$), the aspect ratio ($H_s/W_s$) is $\approx$ 12 in both cases. **(a)** Comparison of contact lines ($W_s = 20d_p$, black) ($W_s = 40d_p$, brown). **(b)** The predications of RW in the larger silo ($W_s = 40d_p$) compared to the estimates from particle dynamics simulations. $F$ is the average contact force at a given depth from the free surface of the silo ($F_n$ in DEM, and $F^{RW}$ in RW (see Methods section for details)), $T_W$ is the total weight of the system.

## 2. Contact model

### *Hertz spring-based particle-particle interaction model*

To study the significance of the type of spring used in the particle-particle interaction model, we simulated narrow source deposited piles with particles interacting by Hertzian spring. In the Hertzian contact model, the normal and tangential forces imparted on $i$ by $j$ are

$$\boldsymbol{F}^{Hertz}_{n_{ij}} = \delta_{ij}^{1/2} R_{eff}^{1/2} \boldsymbol{F}_{n_{ij}} \qquad (9)$$

$$\boldsymbol{F}^{Hertz}_{t_{ij}} = \delta_{ij}^{1/2} R_{eff}^{1/2} \boldsymbol{F}_{t_{ij}} \qquad (10)$$

where $R_{eff} \equiv R_i R_j/(R_i + R_j)$ is the effective radius of the two spheres, and $\boldsymbol{F}_{n_{ij}}$, $\boldsymbol{F}_{t_{ij}}$ are the normal, tangential forces given by equations (5) and (6) respectively, in Supplementary Note 1. As shown in Supplementary Figure 7, the spatial structure and stress distribution are independent of the type of spring used. The parameters used in the simulations are given in Supplementary Table 3. The piles were created by depositing particles from a lower height of deposition, as described in Supplementary Note 2. The dimensions of the piles created are given in Supplementary Table 5.

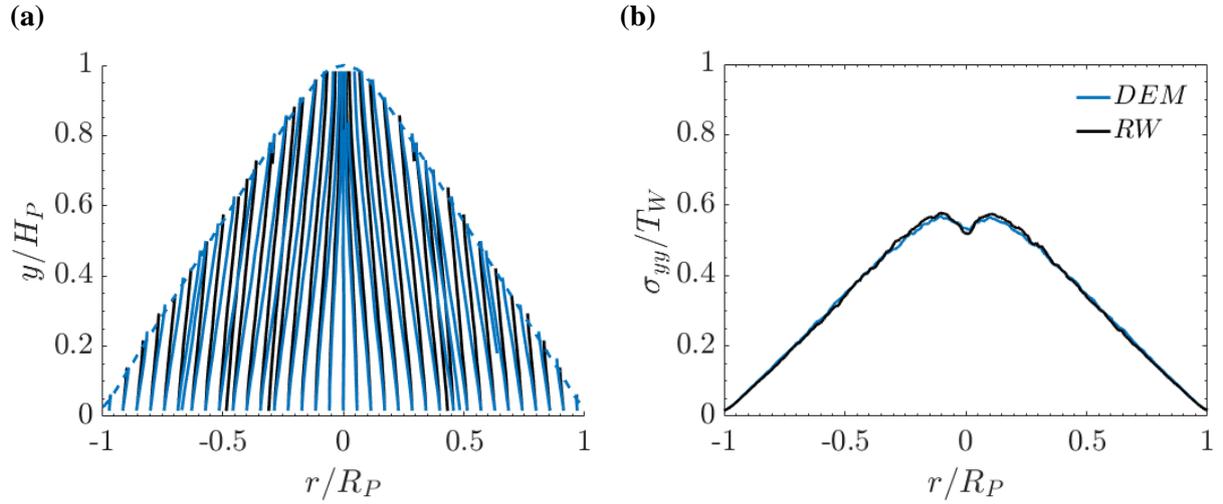

**Supplementary Figure 7:** Spatial structure of piles created by depositing particles interacting with a Hertzian spring in the contact model. Dashed lines represent the average height of the free surface of the pile. **(a)** Contact lines (black) and force lines (blue). **(b)** RW predications compared with the estimates from particle dynamics simulations. $\sigma_{yy}$ is the normal stress measured at the supporting base and $T_W$ is the total weight of the pile.

| Parameter | Value |
|---|---|
| $k_n (N/m)$ | $10^6$ |
| $k_t (N/m)$ | $\frac{2}{7} k_n$ |
| $\gamma_n (1/sec)$ | 8698 |
| $\gamma_t (1/sec)$ | $\frac{1}{2} \gamma_n$ |

**Supplementary Table 3:** Parameters used in the Hertz spring-based contact model.

## 3. Interaction parameters in the linear spring and dashpot model

In this section, we describe the effect of various parameters in the linear spring and dashpot model used in this study.

### 3.1 Spring stiffness

The contact line patterns, and the base stress distribution do not show a considerable difference with varying spring stiffness of the particles (see Supplementary Figure 8). Interestingly with increase in the spring stiffness coefficient of the particles, we find that the predictions of the random walk agree well with the DEM results. This further shows that, the random walk is a reasonable model of force transmission in granular piles, as particles with large spring stiffness values correspond well with realistic granular materials. We note that, the piles made of particles with $k_n = 10^6$ shown in Supplementary Figure 8 were created by depositing particles from a lower height of deposition (see Supplementary Note 2). Here, the value of $\gamma_n$ chosen is such that the normal coefficient of restitution is 0.7. Other interaction parameters are as described in Supplementary Table 4, and $\mu = 0.5$. The dimensions of the piles created are given in Supplementary Table 5.

(a)

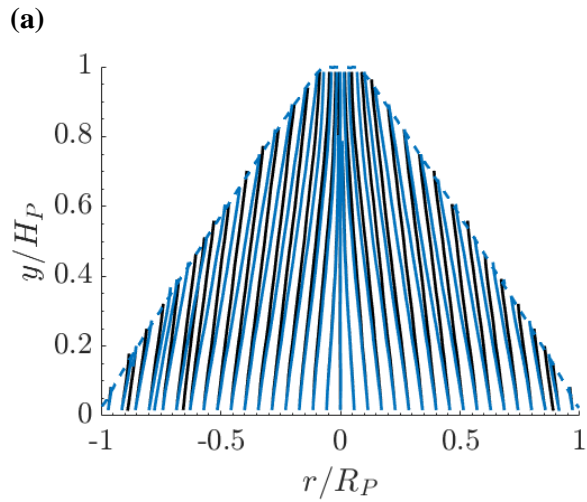

(b)

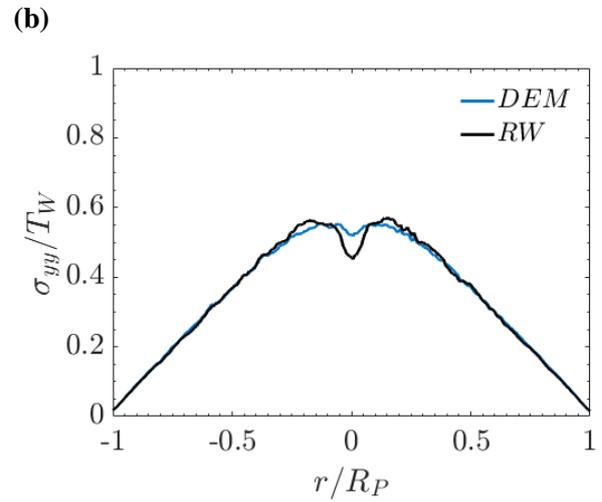

(c)

(d)

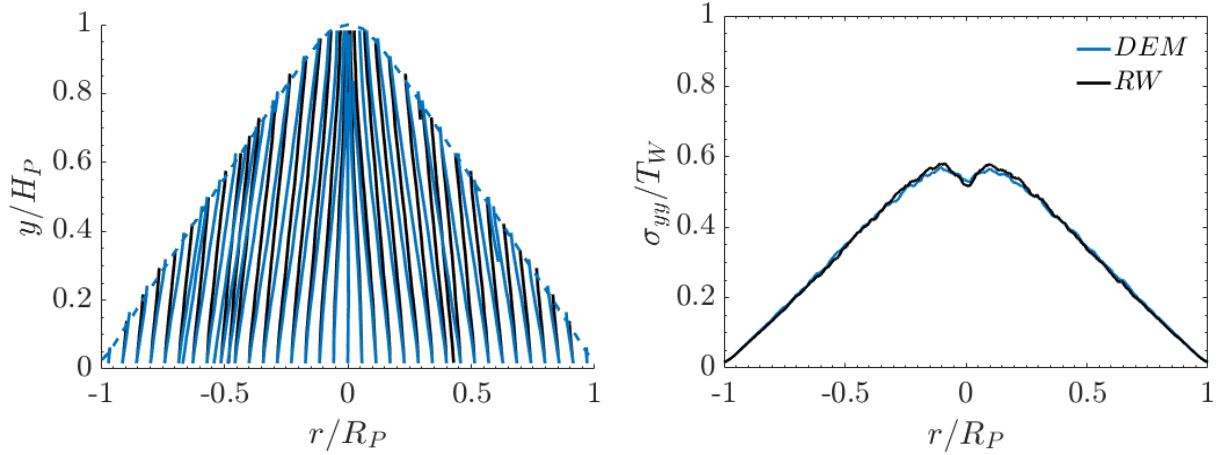

**Supplementary Figure 8:** Effect of spring stiffness on the spatial structure of piles created by depositing particles from a narrow source. **(a)** Contact lines (black) and force lines (blue) of piles made of particles with $k_n = 10^4$. **(b)** RW predications compared with the estimates from particle dynamics simulations of piles made of particles with $k_n = 10^4$. **(c)** Contact lines (black) and force lines (blue) of piles made of particles with $k_n = 10^6$. **(d)** RW predications compared with the estimates from particle dynamics simulations of piles made of particles with $k_n = 10^6$. In both cases, dashed lines represent the average height of the free surface of the pile. $\sigma_{yy}$ is the normal stress measured at the supporting base and $T_W$ is the total weight of the pile.

| Parameter | Case 1 | Case 2 | Case 3 |
|---|---|---|---|
| $k_n (N/m)$ | $10^4$ | $10^5$ | $10^6$ |
| $k_t (N/m)$ | $\frac{2}{7} k_n$ | $\frac{2}{7} k_n$ | $\frac{2}{7} k_n$ |
| $\gamma_n (1/sec)$ | 864 | 2750 | 8698 |
| $\gamma_t (1/sec)$ | $\frac{1}{2}\gamma_n$ | $\frac{1}{2}\gamma_n$ | $\frac{1}{2}\gamma_n$ |

**Supplementary Table 4:** Parameters used in in the linear spring and dashpot model for different values of normal spring stiffness, $k_n$.

### 3.2 Friction coefficient

The coefficient of friction determines the slope of the free surface or angle of repose of the pile. As expected, the angle of repose and consequently slope of the free surface of the piles are different for piles made up of particles with different friction coefficient (Supplementary Table 5). However, we find no significant changes in the contact line profiles or the stress distribution at the base (see Supplementary Figure 9). The dimensions of the piles created are given in Supplementary Table 5. Other interaction parameters are as described in Supplementary Figure 1.

**(a)**                          **(b)**

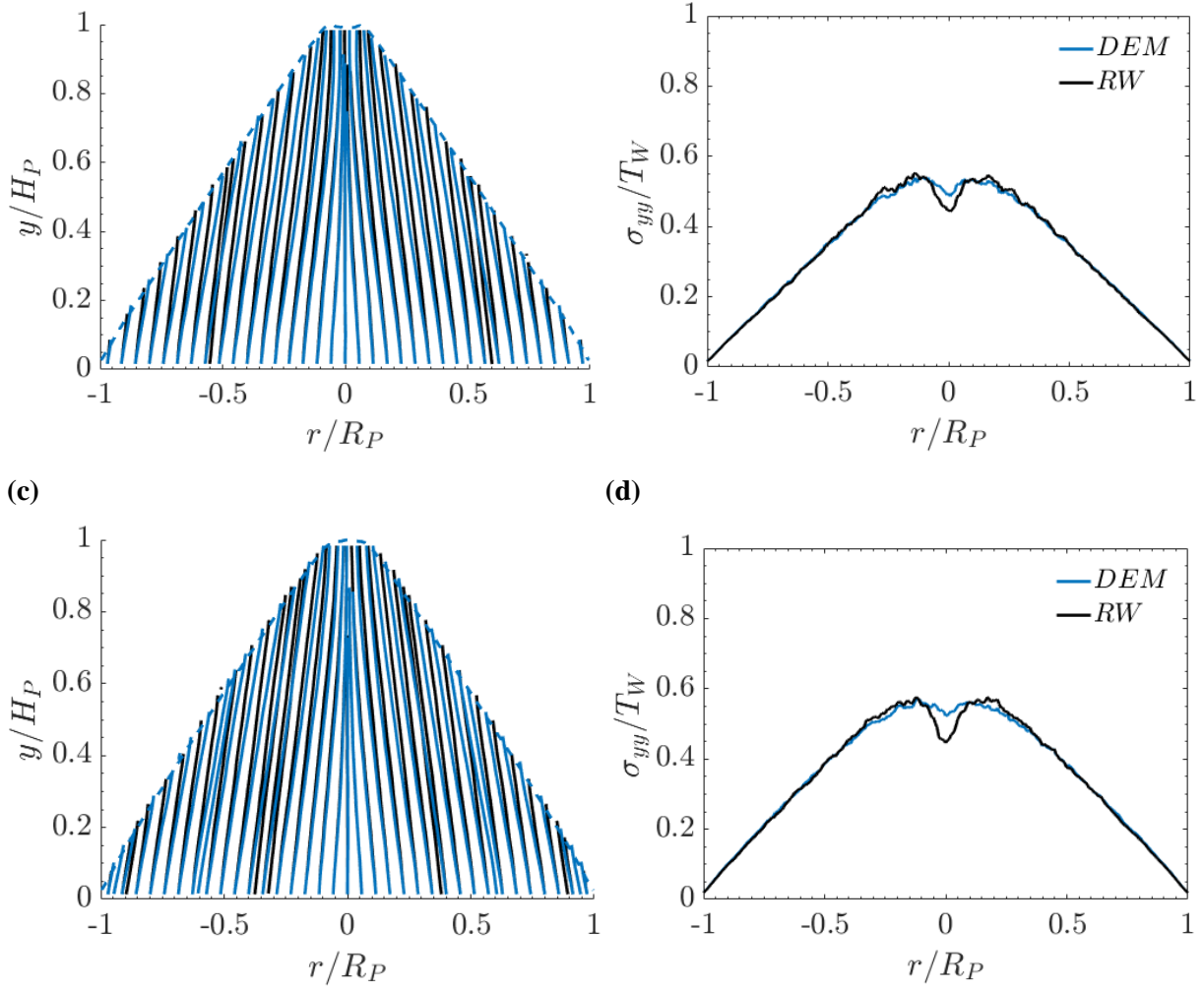

**Supplementary Figure 9:** Effect of friction coefficient on the spatial structure of piles created by depositing particles from a narrow source. **(a)** Contact lines (black) and force lines (blue) of piles made of particles with $\mu = 0.25$. **(b)** RW predications compared with the estimates from particle dynamics simulations of piles made of particles with $\mu = 0.25$. **(c)** Contact lines (black) and force lines (blue) of piles made of particles with $\mu = 0.75$. **(d)** RW predications compared with the estimates from particle dynamics simulations of piles made of particles with $\mu = 0.75$. In both cases, dashed lines represent the average height of the free surface of the pile. $\sigma_{yy}$ is the normal stress measured at the supporting base and $T_W$ is the total weight of the pile.

### 3.3 Size distribution of the particles

A key finding of our study is the importance of particle size distribution on the spatial structure of the pile. A complete understanding of the effect of polydispersity in granular piles requires a detailed study on the type of distribution and the range of polydispersity. Hence, we studied piles made of particles with a narrow distribution of particle sizes compared to the range of polydispersity of 20% used in all polydisperse cases in this study. Here, $\delta = 10\%$, and the particle sizes were uniformly distributed. As shown in Supplementary Figure 10, we find that the contact lines and base stress distributions are similar. The dimensions of the piles created are given in Supplementary Table 5. Other interaction parameters are as described in Supplementary Figure 1, and $\mu = 0.5$.

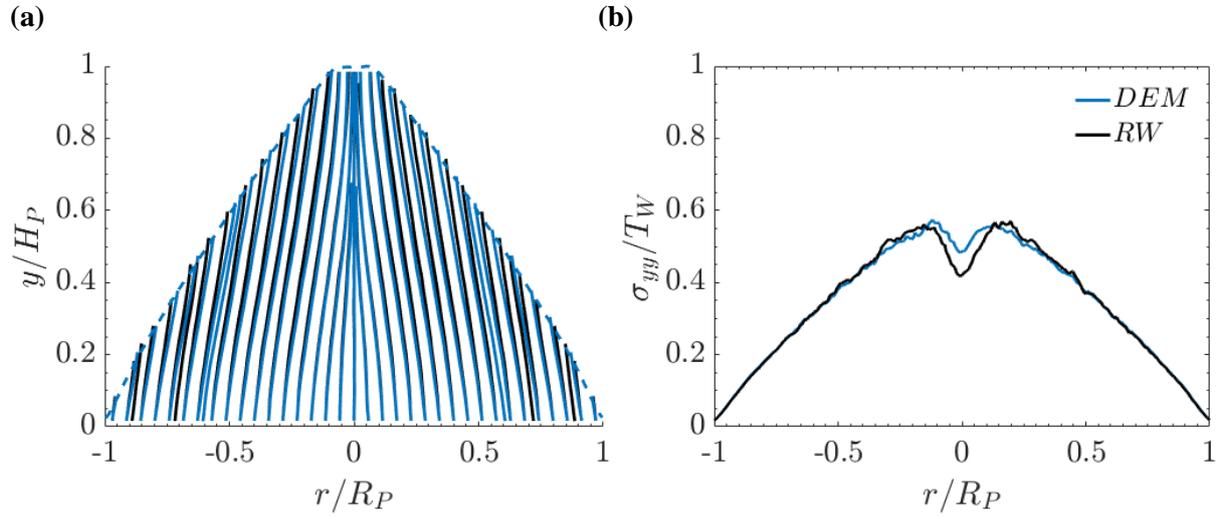

**Supplementary Figure 10:** Spatial structure of piles created by depositing particles with less polydispersity and deposited from a narrow source. **(a)** Contact lines (black) and force lines (blue). Dashed lines represent the average height of the free surface of the pile. **(b)** RW predications compared with the estimates from particle dynamics simulations. $\sigma_{yy}$ is the normal stress measured at the supporting base and $T_W$ is the total weight of the pile.

| Parameters | Height, $H_P$ | Radius, $R_P$ | Angle of repose, $\phi_r$ |
|---|---|---|---|
| Hertzian contact model | 39 | 150 | 14.57 |
| $\mu = 0.25$ | 40 | 152.5 | 14.70 |
| $\mu = 0.5$ | 42 | 144.5 | 16.20 |
| $\mu = 0.75$ | 43 | 140.5 | 17.01 |
| $k_n = 10^4$ | 42 | 145 | 16.15 |
| $k_n = 10^5$ | 42 | 144.5 | 16.20 |
| $k_n = 10^6$ | 39 | 150 | 14.57 |
| $\delta = 10\%$ | 41 | 142.5 | 16.05 |

**Supplementary Table 5:** Dimensions of 2d piles deposited from a narrow source for different values of parameters in the linear spring and dashpot model. In all cases, the radius of the pile $(R_p)$ is the minimum distance from the centre at which the average height of the pile is $\approx 1 d_p \pm (d_p/4)$.

## 4. Averaging procedure and number of independent configurations studied

### 4.1 Base stress profiles in piles

**(a) 2d piles**

The base stress profiles in all cases of 2d piles were obtained by running averages with a window length of $7 d_p$, we found that $7 d_p$ is necessary for obtaining reasonably smooth averages of base stress profiles. Hence, for every base particle of diameter $d_p$ at a horizontal coordinate $x_p$, the base stress values are averaged in the range $[x_p - 3 d_p, x_p + 3 d_p]$. The number of configurations used in obtaining the results reported for 2d piles is 400 for all cases, except for results reported in Fig. 2-4, where 1000 configurations were used.

**(b) 3d piles**

The base stress profiles in 3d piles were obtained by averaging in radial rings of width, $\Delta r = r_o - r_i = 3 d_p$, where $r_i$ is the inner radius and $r_o$ is the outer radius of the ring. The number of configurations used in obtaining the results reported for 3d piles is 30 for all cases.

**(c) 2d and 3d silos**

The normal contact force in all cases of 2d and 3d silo packings were averaged in vertical bins of height, $\Delta h = 5 d_p$. The number of configurations used in obtaining the results reported for 2d and 3d silo packings is 600 and 300 respectively for all cases.

**Supplementary Note 3**

**Comparison of the contact lines, force lines and random walk lines in pile and silo packings**

In all types of packings considered in this study, we find that the contact lines and random walk lines are almost similar (see Supplementary Figure 11). In granular piles, we find that the force lines too agree well with contact lines and random walk lines, except monodisperse case (see Supplementary Figure 12a-e). In silos, the force lines show strong curvature towards the lateral walls compared to the contact lines (see Supplementary Figure 12f-h).

**(a)** 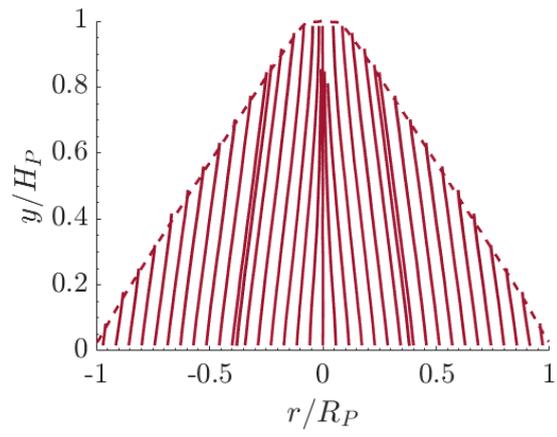 **(b)** 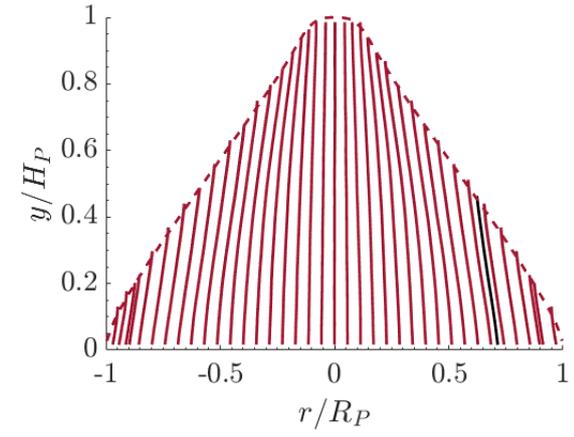

**(c)** 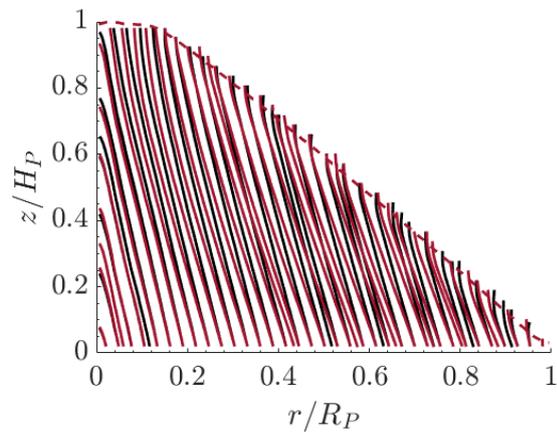 **(d)** 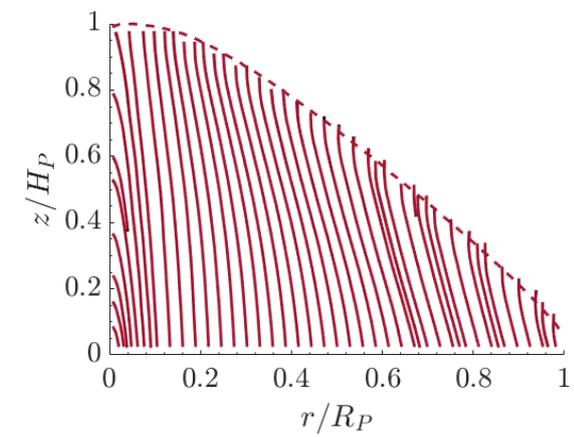

**(e)** 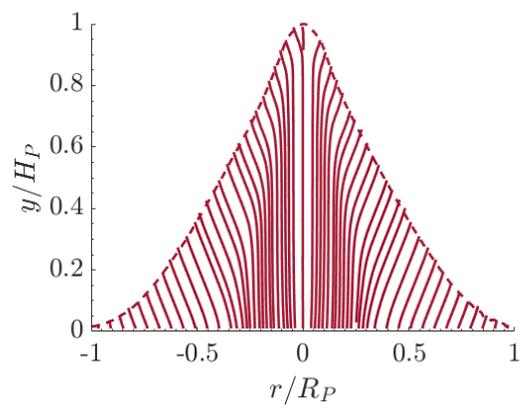

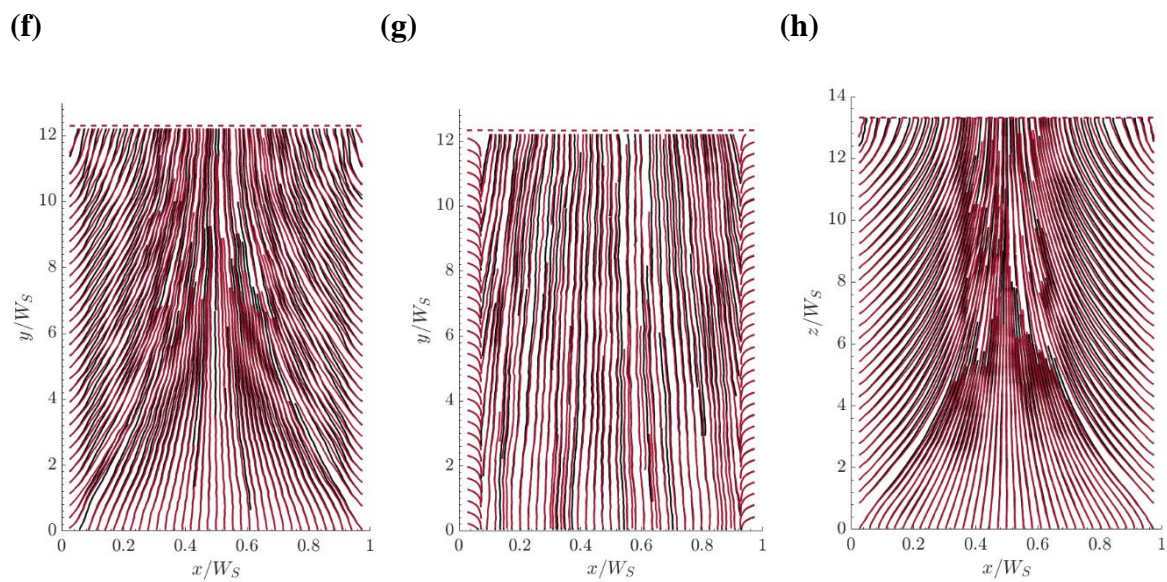

**Supplementary Figure 11:** Comparison on contact lines and random walk lines in pile and silo packings. **(a)** 2d piles deposited from a narrow source. **(b)** 2d piles deposited by raining. **(c)** 3d piles deposited from a narrow source. **(d)** 3d piles deposited by raining. **(e)** 2d piles made up of monodisperse particles deposited from a narrow source. **(f)** 2d silo with rough frictional walls. **(g)** 2d silo with smooth frictionless walls. **(h)** 3d silo with rough frictional walls. In all cases, dashed lines represent the average height of the free surface of the pile.

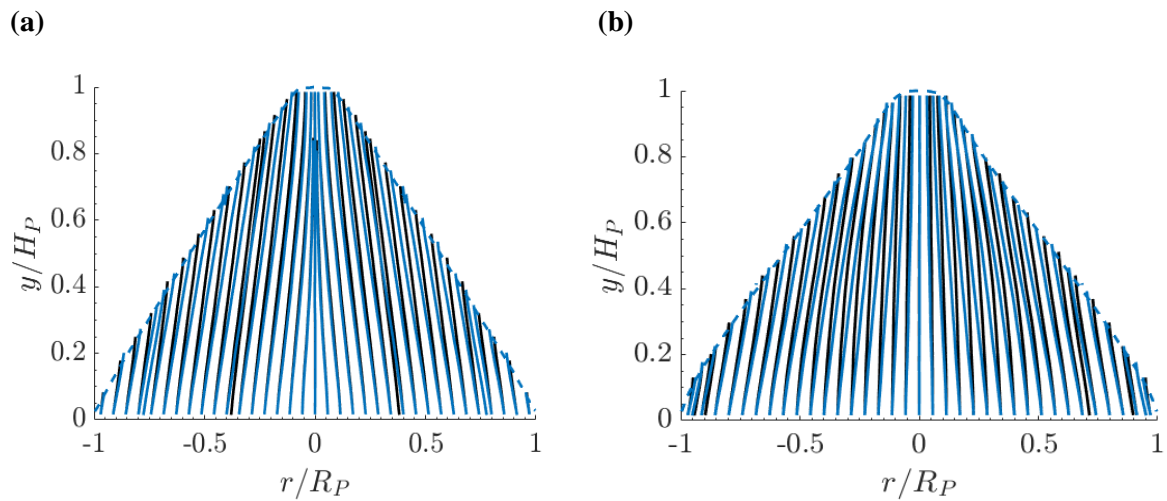

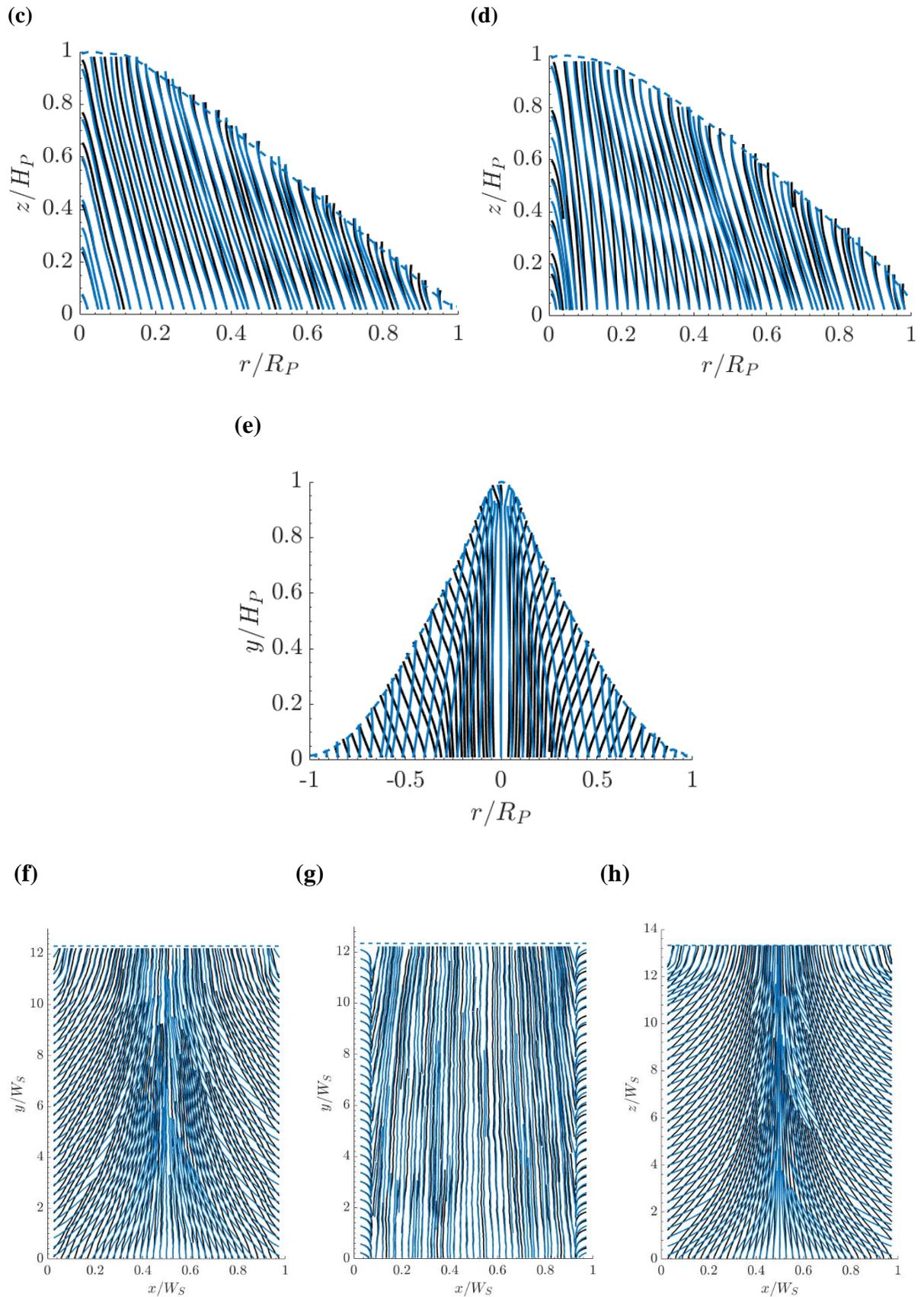

**Supplementary Figure 12:** Comparison on contact lines and force lines in pile and silo packings. **(a)** 2d piles deposited from a narrow source. **(b)** 2d piles deposited by raining. **(c)** 3d piles deposited from a narrow source.

**(d)** 3d piles deposited by raining. **(e)** 2d piles made up of monodisperse particles deposited from a narrow source. **(f)** 2d silo with rough frictional walls. **(g)** 2d silo with smooth frictionless walls. **(h)** 3d silo with rough frictional walls. In all cases, dashed lines represent the average height of the free surface of the pile.

**Supplementary Note 4**

**Details of supplementary videos**

In all the following supplementary videos, for polydisperse packings, the particle sizes were chosen from a uniform distribution with lower and upper limits of $0.8d_p$ and $1.2d_p$ respectively, where $d_p$ is the mean diameter. We refer to the displacement vector field as $\vec{S}$, and is defined as $\vec{r_j} - \vec{r_i}$, where $\vec{r_i}$ is the position vector of the particle at time $t_i$. We have chosen the $\Delta t_{ij} = t_j - t_i$ as $1\backslash 4\ sec$, we found this time interval sufficient for obtaining a smooth visualisation of time evolution of both $\vec{S}$ and $\vec{R}$ vector fields, and in all cases of pile and silo packings.

And, all the properties $(\vec{S}, \vec{R})$ were averaged in a $1 \times 1$ $(d_p \times d_p)$ spatial grid, which was further averaged over 300 independent realizations. Note, here unlike $\vec{R}$, $\vec{S}$ in all directions are considered for obtaining the averages. Also, the time of the video does not correspond to the simulation time and is given below. The video shows the time evolution of all cases till $|\vec{S}|$ decreases to less than $0.01d_p$ in every spatial grid.

*Supplementary videos 1-5 (time evolution of contact lines)*

Evolution of contact lines in 2d pile and silo packings. The lines indicate the streamlines of the spatially averaged $\vec{R}$ vector field. The background colour shows the contour maps based on the particle IDs. The particles IDs are assigned based on the time of deposition scaled by the maximum particle ID (the ID of the last deposited particle).

1. **Supplementary video 1** – Evolution of contact lines in polydisperse piles deposited from a narrow source. The simulation time of the video is 42 seconds.
2. **Supplementary video 2** – Evolution of contact lines in polydisperse piles created by raining the particles. The simulation time of the video is 16 seconds.
3. **Supplementary video 3** – Evolution of contact lines in monodisperse piles deposited from a narrow source. The simulation time of the video is 47 seconds.

4. **Supplementary video 4** – Evolution of contact lines in polydisperse silo packings created by raining the particles into a container with frictional walls. The simulation time of the video is 18 seconds.
5. **Supplementary video 5** – Evolution of contact lines in polydisperse silo packings created by raining the particles into a container with smooth frictionless walls. The simulation time of the video is 46 seconds.

*Supplementary videos 6-10 (time evolution of displacement field of particles)*

Evolution of streamlines based on the displacement field ($\vec{S}$) of particles in 2d pile and silo packings. The lines indicate the streamlines of the spatially averaged displacement field, $\vec{S}$, and the background colour shows the contour map of $|\vec{S}|$ scaled by $d_p$ in logarithmic scale ($log_{10}$). Note that for clarity, the streamlines are shown only if $|\vec{S}| \geq 0.01 d_p$, And, the maximum and minimum value of the background colour is set as $100 d_p$ and $0.01 d_p$ respectively. If $|\vec{S}|$ is greater than $100 d_p$ then it is set to $100 d_p$, and similarly, if $|\vec{S}|$ less than $0.01 d_p$ then it is set to $0.01 d_p$.

1. **Supplementary video 6** – Evolution of streamlines based on $\vec{S}$ in polydisperse piles deposited from a narrow source. The simulation time of the video is 42 seconds.
2. **Supplementary video 7** – Evolution of streamlines based on $\vec{S}$ in polydisperse piles created by raining the particles. The simulation time of the video is 16 seconds.
3. **Supplementary video 8** – Evolution of streamlines based on $\vec{S}$ in monodisperse piles deposited from a narrow source. The simulation time of the video is 47 seconds.
4. **Supplementary video 9** – Evolution of streamlines based on $\vec{S}$ in polydisperse silo packings created by raining the particles into a container with frictional walls. The simulation time of the video is 18 seconds. In frictional silos, a perceptible flow is observed only in the growing free surface region. Hence, we shift the vertical axis accordingly in time to clearly show the flow in the growing free surface region.
5. **Supplementary video 10** – Evolution of streamlines based on $\vec{S}$ in polydisperse silo packings created by raining the particles into a container with smooth frictionless walls. The simulation time of the video is 46 seconds.

**Supplementary References**